\documentclass[preprint,showpacs,preprintnumbers,amsmath,amssymb]{revtex4}

\usepackage{graphicx}
\usepackage{dcolumn}
\usepackage{bm}
\usepackage{amssymb}

\begin{document}
\title{Responses to applied forces and the Jarzynski equality \\
in classical oscillator systems coupled to finite baths: \\
An exactly solvable non-dissipative non-ergodic model
}

\author{Hideo Hasegawa}
\altaffiliation{hideohasegawa@goo.jp}
\affiliation{Department of Physics, Tokyo Gakugei University,  
Koganei, Tokyo 184-8501, Japan}%

\date{\today}

\begin{abstract}
Responses of small open oscillator systems to applied external forces
have been studied with the use of
an exactly solvable classical Caldeira-Leggett (CL) model
in which a harmonic oscillator (system) is coupled to finite $N$-body oscillators (bath) 
with an identical frequency ($\omega_n=\omega_o$ for $n=1$ to $N$).
We have derived exact expressions for positions, momenta and energy of the system
in nonequilibrium states and for work performed by applied forces. 
Detailed study has been made on an analytical method for canonical averages
of physical quantities over the initial equilibrium state, which is much superior 
than numerical averages commonly adopted in simulations of small systems.
The calculated energy of the system which is strongly coupled to finite bath 
is fluctuating but non-dissipative.
It has been shown that the Jarzynski equality (JE) is valid in non-dissipative, 
non-ergodic open oscillator systems regardless of the rate of applied ramp force.

\end{abstract}

        

\maketitle
\newpage
\section{Introduction}

In the last decade, a significant progress has been made in our understanding
of nonequilibrium statistics. 
Experimental and theoretical studies have been developed on small systems
such as quantum dots and biological molecular machines 
which generally operate away from equilibrium
(for reviews, see Refs. \cite{Busta05,Ritort07,Ciliberto10}).
The development of modern techniques of microscopic manipulation
has promoted experimental studies of small systems.
It has become possible to study the response of small systems
to applied external forces.
In parallel theorists have developed the important three theorems:
the Jarzynski equality (JE) \cite{Jarzynski97}, 
the steady-state and transient fluctuation theorems \cite{Evans93,Crooks99,Narayan04},
and Crook's theorem \cite{Crooks99,Narayan04}.
These fluctuation theorems may be applicable to nonequilibrium systems
driven far from the equilibrium states.
In this paper we pay our attention to a remarkable JE given by
\begin{eqnarray}
e^{- \beta \Delta F} &=& \langle e^{-\beta W} \rangle 
= \int dW \:P(W) \:e^{-\beta W},
\label{eq:A1}
\end{eqnarray}
where $W$ denotes a work made in a system when its parameter 
is changed, the bracket $\langle \cdot \rangle$ expresses
the average over the work distribution function (WDF) $P(W)$
of work performed by a prescribed protocol,
$\Delta F$ stands for the free energy difference between the initial and final equilibrium states,
and $\beta$ $(=1/k_B T)$ is the inverse temperature of the initial state.
Equation (\ref{eq:A1}) includes the second law of thermodynamics:
$\langle W \rangle \geq \Delta F$, 
where the equality holds only for the reversible process.
The JE was originally proposed for classical isolated system and open system weakly coupled to baths 
which are described by the Hamiltonian \cite{Jarzynski97} and the stochastic models 
\cite{Jarzynski97b}.
Jarzynski later proved that the JE is valid for strongly coupled open systems \cite{Jarzynski04}.  
A generalization of the JE to quantum systems has been made
in Refs. \cite{Mukamel00}-\cite{Campisi09}. 

A validity of the JE has been confirmed by some experiments 
\cite{Liphardt02,Wang05,Douarche05,Douarche06,Joubaud07,Joubaud07b}.
Liphardt {\it et. al.} \cite{Liphardt02} have determined the free energy required
to unfold a single RNA chain from non-equilibrium work measurements.
Wang {\it et. al.} \cite{Wang05} have considered a colloidal particle pulled through liquid water
by an optical trap.
Douarche {\it et. al.} \cite{Douarche05,Douarche06} have verified the JE 
for a mechanical oscillator that is driven out of equilibrium by an external force. 
By using a torsion pendulum composed of a brass wire,
Joubaud {\it et. al.,} \cite{Joubaud07,Joubaud07b} have experimentally studied 
the JE of the harmonic oscillator in contact with a thermostat
and driven out of equilibrium by an external force. 

Some criticisms, however, have been raised for the validity of the JE
\cite{Cohen04}-\cite{Sung08}.
Cohen and Mauzerall \cite{Cohen04} pointed out that it is difficult 
to define the distribution and the temperature during the irreversible process.
In response to this criticism, Jarzynski \cite{Jarzynski04} has claimed that the JE holds
if the initial state is in the equilibrium state with the definite
temperature \cite{Jarzynski04}.
It has been pointed out that the JE may be violated in ideal gas model 
\cite{Gross05,Tu05,Bena05,Sung07,Sung08} and in a rigid rotator model 
\cite{Sung05a,Bier05,Sung05c}. 
Therefore it is currently an important issue to examine the validity condition of the JE.

Many studies have been reported for harmonic oscillator systems by both 
experimental \cite{Liphardt02,Douarche05,Douarche06,Joubaud07,Joubaud07b,Ciliberto10} and
theoretical methods \cite{Morgado10}-\cite{Hijar10}.
Theoretical analyses have been made for 
oscillators with the use of the Markovian Langevin model 
\cite{Douarche05,Douarche06,Joubaud07,Joubaud07b,Morgado10},
the non-Markovian Langevin model \cite{Zamponi05,Mai07,Speck07,Ohkuma07},
Fokker-Planck equation \cite{Chaudhury08},
and Hamiltonian model \cite{Jarzynski06,Jarzynski08,Dhar05,Chakrabarti08,Hijar10}.
All of these studies have shown that the JE holds in isolated and open oscillators,
assuming dissipative memory kernels or the over-damped models.
This assumption seems reasonable in the situation under which
the relevant experiments \cite{Douarche05,Douarche06,Joubaud07,Joubaud07b} 
have been performed.
Recent theoretical studies, however, have demonstrated that the energy dissipation
is not realized in a small system coupled to finite thermal baths \cite{Plyukhin01,Hasegawa11}.
This is quite different from the case of infinite baths in which dissipation is realized. 
Indeed, it is commonly believed that the dissipation is realized only 
when the system is coupled to {\it infinite} bath (except for chaotic baths)
\cite{Zwanzig01}. Poicar\'{e} recurrence time is finite for finite bath.

It is necessary to make detailed calculations of responses of small systems
to the applied force such as variations of position and energy of the system, 
which have not been reported as far as we are ware of.
The purpose of the present study is twofold: to make detailed study of 
the response to an applied force and to examine the validity of 
the JE in open harmonic oscillator systems in the non-dissipative situation.
We consider the Caldeira-Leggett (CL) Hamiltonian model \cite{Caldeira81,Caldeira83}, 
adopting a {\it single-$\omega$ bath} containing uncoupled $N$-body oscillators 
with an identical frequency: $\omega_n=\omega_o$ for $n=1$ to $N$ 
[Eq. (\ref{eq:D5a})].
The CL model with a single-$\omega$ bath is exactly solvable.
A similar optic-phonon-mode model for bath was adopted
in a different context from the present study \cite{Rise85}.
In the conventional approach, we obtain the Langevin equation
from the CL model, with which its properties are investigated.
In this study, we have 
directly obtained the Laplace-transformed equation of motion of the system.
The energy and work of the system induced by the applied force
are analytically averaged over the canonical distribution of initial 
equilibrium states. 
Our non-dissipative system-plus-bath yields non-ergodic solutions, for which the JE 
will be shown to be valid in contrast with Refs. \cite{Speck07,Ohkuma07}
claiming the importance of the ergodicity.

The paper is organized as follows.
In the next Sec. II, we derive expressions of response of positions, momenta and
system energy induced by an applied ramp force in open oscillator systems,
by using the CL model with the single-$\omega$ bath mentioned above.
We obtain the WDF and the averaged work with which the validity of the JE
have been investigated. Some numerical calculations are presented.
In Sec. III an application of other types of external forces 
to the system is studied. 
We compare our study with the method using the Langevin model
derived from the CL model. 
Sec. IV is devoted to our conclusion.

\section{The adopted model}
\subsection{Equations of motion}

We consider a system of a classical oscillator coupled to 
a bath consisting of $N$-body uncoupled oscillators described by
the CL model \cite{Caldeira81,Caldeira83},
\begin{eqnarray}
H &=& H_S + H_B + H_I,
\label{eq:B1}
\end{eqnarray}
with
\begin{eqnarray}
H_S &=& \frac{P^2}{2M}+ \frac{M \Omega^2 Q^2}{2}- f(t) Q, 
\label{eq:B1a} \\
H_B &=& \sum_{n=1}^N \left( \frac{p_n^2}{2 m}+ \frac{m \omega_n^2 q_n^2}{2} \right), 
\label{eq:B1b} \\
H_I &=& - \sum_{n=1}^N \left( c_n q_n Q - \frac{c_n^2 Q^2}{2 m \omega_n^2} \right),
\label{eq:B1c}
\end{eqnarray}
where $H_S$, $H_B$ and $H_I$ express one-dimensional Hamiltonians of the system, 
bath and interaction, respectively, $M$ ($m$), $\Omega$ ($\omega_n$), $Q$ ($q_n$) 
and $P$ ($p_n$) denote mass, frequency, position and momentum, respectively, 
of the system (bath), $c_n$ the interaction between the system and bath, and $f(t)$
an applied external force.
Equations of motion for $Q$ and $q_n$ are given by
\begin{eqnarray}
M \ddot{Q} &=& -M \Omega^2 Q 
+ \sum_{n=1}^{N} c_n \left(q_n -\frac{c_n Q}{m \omega_n^2}\right)+ f(t),
\label{eq:B2}\\
m \:\ddot{q}_n &=& -m \omega_n^2 q_n + c_n Q.
\label{eq:B3}
\end{eqnarray}

Applying the Laplace transformation to Eqs. (\ref{eq:B2}) and (\ref{eq:B3}),
we obtain
\begin{eqnarray}
M[s^2 \hat{Q}(s) - \dot{Q}(0)-sQ(0)]
&=& - M \Omega^2 \:\hat{Q}(s)
-\sum_{n=1}^N \left( \frac{c_n^2}{m \omega_n^2} \right) \hat{Q}(s)
+ \sum_{n=1}^N c_n \hat{q}_n(s)+ \hat{f}(s), 
\label{eq:D1}\\
m[s^2 \hat{q}_n(s)-\dot{q}_n(0)-s q_n(0)]
&=& - m \omega_n^2 \:\hat{q}_n(s)+c_n \hat{Q}(s),
\label{eq:D2}
\end{eqnarray}
where 
\begin{eqnarray}
\hat{Q}(s) &=& \int_0^{\infty} dt\: e^{-st} \:Q(t),
\label{eq:D3b}
\end{eqnarray}
and similar expressions for $\hat{q}_n(s)$ and $\hat{f}(s)$.
Solving Eq. (\ref{eq:D2}) in terms of $\hat{q}(s)$ and 
substituting it into Eq. (\ref{eq:D1}),
we obtain 
\begin{eqnarray}
\hat{Q}(s) &=& \hat{G}(s)
\left[\dot{Q}(0)+s \:Q(0)
+ \sum_{n=1}^N \frac{c_n[\dot{q}_n(0)+s \:q_n(0)]}{M (s^2+\omega_n^2)}
+ \frac{\hat{f}(s)}{M} \right],
\label{eq:D3}
\end{eqnarray}
where the Green's function $\hat{G}(s)$ is given by 
\begin{eqnarray}
\hat{G}(s) &=& \left(s^2+\Omega^2 
+ \sum_{n=1}^N \frac{c_n^2 s^2}{M m \omega_n^2 (s^2+\omega_n^2)} \right)^{-1}.
\label{eq:D4}
\end{eqnarray}

In order to make analytic calculation feasible, we consider a bath containing
$N$-body uncoupled oscillators with an identical frequency $\omega_o$
and a uniform coupling $c_o$, as given by
\begin{eqnarray}
\omega_n &=& \omega_o, 
\label{eq:D5a}\\
c_n &=& \frac{c_o}{\sqrt{N}} \hspace{1cm}\mbox{for $n=1$ to $N$}.
\label{eq:D5b}
\end{eqnarray}
We have chosen $c_n$ such that it yields a non-divergent result
in the limit of $N \rightarrow \infty$ in Eq. (\ref{eq:D4})
(related discussion being given in Sec. III.B) \cite{Note2}.
With the use of Eqs. (\ref{eq:D5a}) and (\ref{eq:D5b}), 
$\hat{Q}(s)$ becomes
\begin{eqnarray}
\hat{Q}(s) &=& \hat{G}(s)
\left[\frac{P_0}{M}+s \:Q_0
+ \frac{c_o}{M \sqrt{N}(s^2+\omega_o^2)} 
\sum_{n=1}^N \left(\frac{p_{n0}}{m}+s \: q_{n0} \right)+ \frac{\hat{f}(s)}{M} \right],
\label{eq:D6}
\end{eqnarray}
with
\begin{eqnarray}
\hat{G}(s) &=&  \left(s^2+\Omega^2 
+ \frac{c_o^2 s^2}{M m \omega_o^2 (s^2+\omega_o^2)} \right)^{-1},
\label{eq:D7}
\end{eqnarray}
where $P_0=M \dot{Q}(0)$, $Q_0=Q(0)$, $p_{n0}=m \dot{q}_n(0)$ and $q_{n0}=q_n(0)$.
Equation (\ref{eq:D7}) may be rewritten as
\begin{eqnarray}
\hat{G}(s) 
&=& \frac{s^2+\omega_o^2}{[(s^2+\Omega^2)(s^2+\omega_o^2)+c_o^2 s^2/M m \omega_o^2]}, \\
&=& \sum_{i=1}^2 \frac{b_i}{(s^2+a_i^2)},
\label{eq:D8}
\end{eqnarray}
with
\begin{eqnarray}
a_i^2 &=& \frac{1}{2}\left[\Omega^2+\omega_o^2+\frac{c^2}{M m \omega_o^2} 
+ (-1)^{i-1} \sqrt{D_o} \right] 
\hspace{1cm} \mbox{($i=1,2$)}, 
\label{eq:D9}\\
D_o &=& (\Omega^2-\omega_o^2)^2 + \frac{2c_o^2(\Omega^2+\omega_o^2)}{M m \omega_o^2}
+\frac{c_o^4}{M^2 m^2 \omega_o^4} \geq 0, 
\label{eq:D10}\\
b_1 &=& \frac{a_1^2-\omega_o^2}{a_1^2-a_2^2},\;\;\;\;
b_2 = \frac{\omega_o^2-a_2^2}{a_1^2-a_2^2}.
\label{eq:D11}
\end{eqnarray}
Then Eq. (\ref{eq:D6}) becomes
\begin{eqnarray}
\hat{Q}(s) = \sum_{i=1}^2 \frac{b_i}{(s^2+a_i^2)}
\left[ \frac{P_0}{M}+ s Q_0 + \frac{c_o}{M \sqrt{N} (s^2+\omega_o^2)}
\sum_{n=1}^N \left(\frac{p_{n0}}{m}+s \:q_{n0}   \right) + \frac{\hat{f}(s)}{M} \right],
\label{eq:D12b}
\end{eqnarray}
whose inverse Laplace transformation yields
\begin{eqnarray}
Q(t) &=& \Phi(t)+X_Q(t) Q_0+X_P(t) P_0 
+Y_q(t) \sum_{n=1}^N q_{n0}+Y_p(t) \sum_{n=1}^N p_{n0}, 
\label{eq:D13}
\end{eqnarray}
with
\begin{eqnarray}
\Phi(t) &=& \sum_{i=1}^2 \frac{b_i}{M a_i} \int_{0}^t \sin a_i (t-t') f(t')\:dt', 
\label{eq:D12}\\
X_Q(t) &=& \sum_{i=1}^2 b_i \cos a_i t, 
\label{eq:D12f}\\
X_P(t) &=& \sum_{i=1}^2 \left( \frac{ b_i}{M a_i} \right) \sin a_i t, 
\label{eq:D12g}\\
Y_q(t) &=& \sum_{i=1}^2  \left( \frac{b_i \:c_o}{M \sqrt{N}} \right) 
\frac{(\cos \omega_o t-\cos a_i t)}{(a_i^2-\omega_o^2)}, 
\label{eq:D12e} \\
Y_p(t) &=& \sum_{i=1}^2  \left( \frac{b_i \:c_o}{M \sqrt{N}} \right) 
\frac{(a_i \sin \omega_o t-\omega_o \sin a_i t)}
{m \omega_o  a_i (a_i^2-\omega_o^2)}.
\label{eq:D12d}
\end{eqnarray}

\subsection{Position, momentum and system energy}

It is necessary to evaluate physical quantities averaged over the canonical distribution 
of initial states, $Q_0$, $P_0$, $\{q_{n0} \}$ and $\{p_{n0} \}$, 
of the equilibrium coupled system-and-bath $H(t=0)$. 
In order to make such evaluations, we need following 
(fluctuation-dissipation) relations for $f(0)=0$ given by
\begin{eqnarray}
M \Omega^2 \langle Q_0^2 \rangle_0 &=& \frac{ \left<P_0^2 \right>_0}{M}
= k_B T =\frac{1}{\beta}, 
\label{eq:H2a}\\
m \omega_o^2 \langle q_{n0} \:q_{\ell0} \rangle_0 
&=& k_B T\:\delta_{n\ell} + \frac{c_n c_{\ell} k_B T}{m \omega_o^2 M \Omega^2}, 
\label{eq:H2c}\\
\frac{\left< p_{n0}\:p_{\ell0} \right>_0}{m} &=& k_B T \:\delta_{n\ell}, \\
\langle Q_0 \:q_{n0} \rangle_0 &=& \frac{c_n k_B T}{m \omega_o^2 M \Omega^2},
\label{eq:H2d}\\
\langle P_0 \:Q_0 \rangle_0 &=& \langle P_0 \:q_{n0} \rangle_0= \langle P_0 \:p_{n0} \rangle_0 
= \langle p_{n0} \:q_{\ell 0} \rangle_0
= \langle p_{n0} \:Q \rangle_0=0,\\
\langle Q_0 \rangle_0 &=& \langle P_0 \rangle_0
= \langle q_{n0} \rangle_0 = \langle p_{n0} \rangle_0 =0,
\label{eq:H2b}
\end{eqnarray}
with
\begin{eqnarray}
\langle O \rangle_0
&\equiv & \frac{{\rm Tr} \:\{e^{-\beta H(0)} \:O\} }{{\rm Tr} \:e^{-\beta H(0)}},
\label{eq:H3b} 
\end{eqnarray}
where $O$ denotes an operator
and ${\rm Tr}$ the trace over initial state of $H(0)$ with $Q_0$, $P_0$, 
$\{ q_{n0} \}$ and $\{p_{n0} \}$.
Equations (\ref{eq:H2c}) and (\ref{eq:H2d}) arise from the relation,
\begin{eqnarray}
m \omega_o^2 \left< \left(q_{n0}-\frac{c_n Q_0}{m \omega_o^2} \right) 
\left(q_{\ell 0}-\frac{c_{\ell} Q_0}{m \omega_o^2} \right) \right>_0
&=& k_B T \:\delta_{n\ell}.
\end{eqnarray}
In the limit of $c_o=0$,  Eqs. (\ref{eq:H2a})-(\ref{eq:H2b}) reduce to the well-known result
for isolated system and bath.
With the use of Eqs. (\ref{eq:D13}) and (\ref{eq:H2b}), 
the averaged position and momentum of the system are given by
\begin{eqnarray}
\bar{Q}(t) & = & \langle Q(t) \rangle_0 =\Phi(t)
= \int_{0}^t \:\chi(t-t') f(t')\:dt',
\label{eq:H4} \\
\bar{P}(t) & = & \langle P(t) \rangle_0 = M \dot{\Phi}(t)
=  M \int_{0}^t \: \dot{\chi}(t-t') f(t')\:dt',
\label{eq:H5}
\end{eqnarray}
with the time-dependent susceptibility $\chi(t)$,
\begin{eqnarray}
\chi(t) = \sum_{i=1}^2 \frac{b_i \sin a_i t}{M a_i},
\end{eqnarray}
where dot $(\cdot)$ stands for the derivative with respect to time.
It is easy to see from Eq. (\ref{eq:D6}) 
that the Laplace-transformed susceptibility
is given by $\hat{\chi}(s)=\hat{G}(s)/M$.
The frequency-dependent susceptibility $\chi(\omega)$ is given by
\begin{eqnarray}
\chi(\omega) &=& \int_{0}^{\infty} e^{- i \omega t} \chi(t)\:dt
=\hat{\chi}(- i \omega), \\ 
&=& - \frac{1}{M} \sum_{i=1}^2 \frac{b_i}{a_i (\omega^2-a_i^2)},
\end{eqnarray}
whose imaginary part becomes
\begin{eqnarray}
{\rm Im} \;\chi(\omega) &=& \sum_{i=1}^2 
\left( \frac{\pi M b_i}{2 a_i} \right)
[\delta(\omega-a_i) -\delta(\omega+a_i)].
\end{eqnarray}

The system energy $\bar{E}_S$ averaged over the initial state is given by 
\cite{Note1,Hanggi08,Ingold09,Gelin09}
\begin{eqnarray}
\bar{E}_S &=& \langle E_S \rangle_0
= \frac{M}{2} \left< \dot{Q}^2 \right>_0
+\frac{M \Omega^2}{2}  \left< Q^2 \right>_0  -f(t) \left< Q \right>_0 . 
\label{eq:H1}
\end{eqnarray}
By using (\ref{eq:H2a})-(\ref{eq:H2b}) and  Eqs. (\ref{eq:H1}), 
we obtain $\bar{E}_S$ given by
\begin{eqnarray}
\bar{E}_S &=& \bar{E}_S^{(0)}+\bar{E}_S^{(f)},
\label{eq:H3}
\end{eqnarray}
with
\begin{eqnarray}
\bar{E}_S^{(0)}  
&=&  \frac{k_B T}{2 M \Omega^2}
\left[M \dot{X}_Q(t)^2+ M \Omega^2 X_Q(t)^2 \right] 
 \nonumber \\
&+&  \frac{M k_B T}{2} \left[M \dot{X}_P(t)^2+ M \Omega^2 X_P(t)^2 \right]
+ \frac{N m k_B T}{2} \left[M \dot{Y}_p(t)^2+ M \Omega^2 Y_p(t)^2 \right] 
\nonumber \\
&+& \frac{N k_B T}{2m \omega^2}
\left(1 + \frac{c_o^2}{m \omega_o^2 M \Omega^2}\right)
\left[M \dot{Y}_q(t)^2+ M \Omega^2 Y_q(t)^2 \right] 
\nonumber \\
&+& \frac{ \sqrt{N} c_o k_B T}{m \omega_o^2 M \Omega^2}
\left[ M \dot{X}_Q(t) \dot{Y}_q(t) + M \Omega^2 X_Q(t) Y_q(t) \right], 
\label{eq:H3a}\\
\bar{E}_S^{(f)} 
&=& \frac{1}{2} \left[M \dot{\Phi}(t)^2 + M \Omega^2 \Phi(t)^2  \right]
- f(t) \Phi(t).
\label{eq:H3c}
\end{eqnarray}
Here $\bar{E}_S^{(0)}$ expresses the system energy depending on the temperature
but independent of the applied force:
$\bar{E}_S^{(f)}$ denotes the response to the force:
$\Phi(t)$, $X_Q(t)$, $X_P(t)$, $Y_q(t)$ and $Y_p(t)$ are given
by Eqs. (\ref{eq:D12})-(\ref{eq:D12d}):
$\dot{X}_Q(t)$, $\dot{X}_P(t)$, $\dot{Y}_q(t)$ and $\dot{Y}_p(t)$
are their derivatives with respect to time.
It is noted that $\bar{Q}(t)$ and $\bar{P}(t)$ 
are independent of $N$ because of the $N$-independent $\Phi(t)$ in Eq. (\ref{eq:D12}).
Furthermore $\bar{E}_S$ does not depend on $N$ because the $N$ factor in the fourth term
of Eq. (\ref{eq:H3a}) is canceled out by the $1/N$ term in $Y_q(t)^2$ in Eq. (\ref{eq:D12e}) 
and because the $\sqrt{N}$ term of the last term of Eq. (\ref{eq:H3a}) is cancelled out by
the $1/\sqrt{N}$ of $Y_q(t)$.  These properties arise from our adopted model 
with $c_n=c_o/\sqrt{N}$ in Eq. (\ref{eq:D5b}) \cite{Note2}.

The advantage of expressions given by Eqs. (\ref{eq:H4}), (\ref{eq:H5}) 
and (\ref{eq:H3})-(\ref{eq:H3c})
is that canonical averages over the initial state have been analytically made and they are
free from the numerical averaging which is one of difficulties in direct simulations
of small systems \cite{Plyukhin01,Hasegawa11,Rosa08,Smith08,Wei09}.

We have so far not specified the form of an external force $f(t)$.
For a while we consider a ramp force given by
\begin{eqnarray}
f(t) &=& \left\{ \begin{array}{ll}
0
\quad & \mbox{for $t < 0 $}, \\ 
g \:(\frac{t}{\tau})
\quad & \mbox{for $0 \leq t < \tau $}, \\ 
g
\quad & \mbox{for $t \geq \tau $},
\end{array} \right. 
\label{eq:B4}
\end{eqnarray}
where 
$\tau$ stands for a duration of
the applied force and $g$ the magnitude of the force at $t \geq \tau$.
For the ramp force, Eq. (\ref{eq:D12}) leads to
\begin{eqnarray}
\Phi(t) &=& \sum_{i=1}^2 \left( \frac{g b_i}{M a_i^3 \tau} \right) (a_i t -\sin a_i t)
\hspace{1cm} \mbox{for $0 \leq t < \tau$}, 
\label{eq:B6}\\
&=& \sum_{i=1}^2 \left(\frac{g b_i}{M a_i^2} \right)
\left( \frac{1}{a_i \tau} \left[a_i \tau +\sin a_i(t-\tau) -\sin a_i t \right]  \right)
\hspace{1cm} \mbox{for $t \geq \tau$}.
\label{eq:B7}
\end{eqnarray}
In the following, we examine the three cases of (1) no couplings $(c_o=0)$, (2) 
transient force $(\tau=0)$ and (3) quasi-static force $(\tau \rightarrow \infty)$.

(1) In the case of $c_o=0$ where Eqs. (\ref{eq:D9})-(\ref{eq:D11}) lead to
$a_1=\Omega$, $a_2=\omega_o$, $b_1=1$ and $b_2=0$,
we obtain
\begin{eqnarray}
\Phi(t) &=&  \left( \frac{g}{M \Omega^3 \tau} \right) (\Omega t -\sin \Omega t)
\hspace{1cm} \mbox{for $0 \leq t < \tau$}, 
\label{eq:C3b}\\
&=& \left(\frac{g}{M \Omega^2} \right)
\left( \frac{1}{\Omega \tau} \left[ \Omega \tau +\sin \Omega(t-\tau)-\sin \Omega t
\right] \right)
\hspace{1cm} \mbox{for $t \geq \tau$}.
\end{eqnarray}
Equations (\ref{eq:D12f}) and (\ref{eq:D12g}) lead to
\begin{eqnarray}
X_Q(t) &=& \cos \Omega t, \\
X_P(t) &=& \left( \frac{1}{M \Omega} \right) \sin \Omega t.
\end{eqnarray}
$\bar{E}_S(t)$ becomes
\begin{eqnarray}
\bar{E}_S(t) &=& k_B T - \frac{g^2}{2 M \Omega^2}
\left[ \left(\frac{t}{\tau}\right)^2- \frac{2(1- \cos \Omega t)}{\Omega^2 \tau^2} \right]
\hspace{1cm}\mbox{for $0 \leq t < \tau$},\\
&=& k_B T - \frac{g^2}{2 M \Omega^2}
\left[ 1 - \frac{2(1- \cos \Omega \tau)}{\Omega^2 \tau^2} \right]
\hspace{1cm}\mbox{for $ t \geq \tau$}.
\end{eqnarray} 
 
(2) In the case of $\tau=0$, Eq. (\ref{eq:B7}) yields
\begin{eqnarray}
\Phi(t) &=& \sum_{i=1}^2 \left( \frac{g b_i}{M a_i^2} \right)
(1-\cos a_i t) 
\hspace{1cm}\mbox{for $t \geq 0$},
\end{eqnarray}
which becomes for $c_o=0$, 
\begin{eqnarray}
\Phi(t) &=&  \left( \frac{g}{M \Omega^2} \right)
(1-\cos \Omega t) 
\hspace{1cm}\mbox{for $t \geq 0$},
\end{eqnarray}
yielding
\begin{eqnarray}
\bar{E}_S(t) &=& k_B T
\hspace{1cm}\mbox{for $t \geq 0$}.
\end{eqnarray} 
 
(3) In the case of $\tau \rightarrow \infty$, Eq. (\ref{eq:B6}) yields
\begin{eqnarray}
\Phi(t) &=& \sum_{i=1}^2  \frac{g b_i}{M a_i^2} \left(\frac{t}{\tau}\right)
\hspace{1cm}\mbox{for $0 \leq t < \infty$},
\end{eqnarray}
which becomes for $c_o=0$, 
\begin{eqnarray}
\Phi(t) &=&  \frac{g}{M \Omega^2} \left(\frac{t}{\tau} \right)
\hspace{1cm}\mbox{for $0 \leq t < \infty$},
\end{eqnarray}
leading to
\begin{eqnarray}
\bar{E}_S(t) &=& k_B T -\frac{g^2}{2 M \Omega^2} \left(\frac{t}{\tau}\right)^2, \\
&=& k_B T -\frac{g^2}{2 M \Omega^2}
\hspace{1cm}\mbox{for $t = \tau \rightarrow \infty$}.
\label{eq:J1}
\end{eqnarray}

We have performed numerical calculations for averaged position, momentum and 
energy of the system with $M=m=1.0$, $\Omega=\omega_o=1.0$ and $g=1.0$ which
are adopted in all our calculations otherwise noticed. 
Position, momentum and energy (work) are measured in units of
$\sqrt{k_B T/M \Omega^2}$, $\sqrt{M k_B T}$ and $k_B T$, respectively.
Model calculations of averaged positions and momenta
are presented in Figs. \ref{fig1}(a)-(h) where solid and dashed curves express
$\bar{Q}(t)$ and $\bar{P}(t)$, respectively.
Figures \ref{fig1}(a) and (b) show the results of $c_o=0.0$ and $c_o=1.0$, respectively,
when a ramp force with $\tau=100$ is applied. Figure \ref{fig1}(a) shows that
$\bar{Q}(t)$ is linearly increased at $0 \leq t < 100.0$, and
it becomes constant at $t \geq 100.0$ where a force $g$ is still applied. 
This behavior is not changed even when
the system-bath coupling is introduced as shown in Fig. \ref{fig1}(b).
Figures \ref{fig1}(c), (e) and (g) show 
$\bar{Q}(t)$ and $\bar{P}(t)$ for
ramp forces with $\tau=10.0$, $5.0$ and 0.0, respectively, applied 
to uncoupled systems ($c_o=0.0$), where regular oscillations are induced. 
Figures \ref{fig1}(d), (f) and (h), however, show that
irregular oscillations are induced by external forces 
with $\tau=10.0$, $5.0$ and 0.0 in coupled systems.

Model calculations of system energy $\bar{E}_S(t)$ are plotted in Figs. \ref{fig2}(a)-(j).
Figures \ref{fig2}(a) and (b) show $\bar{E}_S(t)$ for $c_o=0.0$ and $c_o=1.0$, respectively, 
without external forces [$f(t)=g=0.0$] for which $\bar{E}_S$ is constant.
Figures \ref{fig2}(c) and (e) (Figs. \ref{fig2}(d) and (f)) show $\bar{E}_S$ 
for $c_o=0.0$ ($c_o=1.0$), with applied forces of $\tau=100.0$ and 10.0, respectively, where
$\bar{E}_S$ is gradually decreased by an applied force.
As far as the uncoupled system is concerned, this behavior is not
modified when the force with smaller $\tau$ is applied, as shown by
Figs. \ref{fig2}(g) and (i) for $\tau=5.0$ and $\tau=0.0$, respectively.
However, when the ramp force with smaller $\tau$ is applied to
coupled systems, the behavior is changed: irregular oscillations
are induced in $\bar{E}_S$ as shown by Fig. \ref{fig2}(h) and (j)
for $\tau=5.0$ and $\tau=0.0$, respectively.
These oscillations in coupled systems are realized for ramp forces
with $\tau \lesssim T_o$ where $T_o$ ($= 2 \pi/\Omega$) denotes the
period of system oscillation. 
We note in Figs. \ref{fig2}(h) and (j) that this irregular oscillation is not dissipate,
which has been confirmed with calculations for $t \in [0, 10000]$ (relevant results not shown).
The averaged system energy in the coupled small systems shows irregular 
non-dissipative oscillations although the total energy of the system-plus-bath is constant
\cite{Plyukhin01,Hasegawa11}.

\subsection{Work and work distribution function}

Next we consider a work performed by an applied external force.
By using $Q(t)$ given by Eq. (\ref{eq:D13}), we obtain the work performed 
by the force $f(t)$ applied for $0 \leq t < \tau$ \cite{Jarzynski97},
\begin{eqnarray}
W_0 &=& - \int_0^{\tau} dt\: \dot{f}(t) Q(t), \\
&=& \phi + C_Q Q_0+ C_P P_0+ D_q \sum_{n=1}^N q_{n0}+ D_p \sum_{n=1}^N p_{n0},
\label{eq:D14}
\end{eqnarray}
where
\begin{eqnarray}
\phi &=& - \int_0^{\tau} dt\: \dot{f}(t) \Phi(t),
\label{eq:D14b} \\
C_{\xi} &=& - \int_0^{\tau} dt\: \dot{f}(t) X_{\xi}(t)
\hspace{0.5cm}\mbox{(for $\xi=Q$ and $P$)},
\label{eq:D14c} \\
D_{\eta} &=& - \int_0^{\tau} dt\: \dot{f}(t) Y_{\eta}(t)
\hspace{0.5cm}\mbox{(for $\eta=q$ and $p$)}.
\label{eq:D14d}
\end{eqnarray}
With the use of Eqs. (\ref{eq:D14})-(\ref{eq:D14d}), 
the WDF of $P(W)$ is given by
\begin{eqnarray}
P(W) &=& \left< \delta \left( W- W_0 \right) \right>_0, \\
&=& \frac{1}{2 \pi} \int du \:\exp(iu W)
\left< \exp(- i u W_0)\right>_0,
\label{eq:D20}
\end{eqnarray}
where
\begin{eqnarray}
\left<\exp(-i u W_0) \right>_0
&=&  \exp(- i u \phi) \left( \frac{\beta \Omega}{2 \pi} \right)
\left( \frac{\beta \omega_o}{2 \pi} \right)^N
\int d Q_0 \exp\left[ -\frac{\beta M \Omega^2 Q_0^2}{2} - i u C_Q Q_0\right]
\nonumber \\
&\times& \int d P_0 \exp\left[ -\frac{\beta P_0^2}{2 M} - i u C_P P_0 \right] 
\nonumber \\
&\times& \prod_{n=1}^N \int d q_{n0} 
\exp\left[ -\frac{\beta m \omega_n^2}{2} \left( q_{n0}-\frac{c_n Q_0}{m \omega_o^2} \right)^2
- i u D_q \left(q_{n0}-\frac{c_n Q_0}{m \omega_o^2} \right)
- \frac{i u D_q c_n Q_0}{m \omega_o^2} \right] 
\nonumber \\
&\times&  \prod_{n=1}^N \int d p_{n0} 
\exp\left( -\frac{\beta p_{n0}^2}{2 m} - i u D_p p_{n0} \right).
\end{eqnarray}
Performing the Gauss integrals, we obtain
\begin{eqnarray}
\left<\exp(-i u W_0) \right>_0
&=& \exp\left[-i u \phi -\frac{u^2}{2 \sigma^2} \right], 
\label{eq:D21}
\end{eqnarray}
where
\begin{eqnarray}
\sigma^2 &=& \frac{1}{\beta}\left[ \frac{1}{M \Omega^2}
\left(C_Q+ \frac{\sqrt{N}\:c_o D_q}{m \omega_o^2} \right)^2 +M C_P^2
+ \frac{N D_q^2}{m \omega_o^2} + m N D_p^2 \right].
\label{eq:D22}
\end{eqnarray}

With the use of Eqs. (\ref{eq:D20}) and (\ref{eq:D21}), $P(W)$ is finally given by 
\begin{eqnarray}
P(W) 
&=& \frac{1}{\sqrt{2 \pi \sigma^2}} \exp\left[ - \frac{(W-\mu)^2}{2 \sigma^2} \right],
\label{eq:D23}
\end{eqnarray}
with
\begin{eqnarray}
\mu=\langle W \rangle= \phi,
\label{eq:D24}
\end{eqnarray}
where $\phi$ and $\sigma^2$ are given by Eqs. (\ref{eq:D14b})
and (\ref{eq:D22}), respectively. 
The average of $e^{-\beta W}$ over $P(W)$ is given by
\begin{eqnarray}
\left< e^{-\beta W} \right> 
&=& \int dW \: P(W)\: e^{-\beta W}
= e^{-\beta (\mu-\beta \sigma^2/2)}, 
\label{eq:F3}
\end{eqnarray}
which leads to
\begin{eqnarray}
R &\equiv & -\frac{1}{\beta}\ln \left<e^{-\beta W}\right>, 
\label{eq:F4b} \\
&=& \mu -\frac{\beta \sigma^2}{2}=\phi -\frac{\beta \sigma^2}{2}.
\label{eq:F4}
\end{eqnarray}

It is worthwhile to point out that Eq. (\ref{eq:F3}) may be alternatively obtainable by 
\begin{eqnarray}
\langle e^{-\beta W_0} \rangle_0
&=& e^{- \beta \left(\phi- \beta \sigma^2/2 \right)},
\label{eq:F9}
\end{eqnarray}
where 
$\phi$ and $\sigma^2$ are given by Eqs. (\ref{eq:D14b})
and (\ref{eq:D22}), respectively.

For a ramp force given by Eq. (\ref{eq:B4}), Eqs. (\ref{eq:D14b})-(\ref{eq:D14d}) 
are given by
\begin{eqnarray}
%
\phi &=& - \left(\frac{g^2}{M}\right) \sum_{i=1}^2 b_i \:\left[ \frac{1}{2 a_i^2}
- \frac{(1-\cos a_i \tau)}{a_i^4 \tau^2} \right],
\label{eq:D15}\\
C_Q &=& -g \sum_{i=1}^2 \frac{b_i \sin a_i \tau}{a_i \tau}, 
\label{eq:D16} \\
C_P &=& -\left( \frac{g}{M} \right) 
\sum_{i=1}^2 \frac{b_i (1- \cos a_i \tau)}{a_i^2 \tau}, 
\label{eq:D17}\\
D_q &=& -\left( \frac{c_o g}{\sqrt{N} M} \right)
\sum_{i=1}^2 \frac{b_i ( a_i \sin \omega_o \tau-\omega_o \sin a_i \tau)}
{a_i  \omega_o \tau(a_i^2-\omega_o^2)},
\label{eq:D18}\\
D_p &=&-\left( \frac{c_o g}{\sqrt{N} M m} \right)
\sum_{i=1}^2 \frac{b_i [a_i^2(1-\cos \omega_o \tau)-\omega_o^2(1- \cos a_i \tau)]}
{a_i^2  \omega_o^2 \tau (a_i^2-\omega_o^2)}.
\label{eq:D19}
\end{eqnarray}

It is noted that $R$ given by Eqs. (\ref{eq:D22}), (\ref{eq:F4}), (\ref{eq:D15})-(\ref{eq:D19}) 
is independent of $N$ because $\sqrt{N}$ factor in the first term of 
Eq. (\ref{eq:D22}) is cancelled out by $1/\sqrt{N}$ of $D_q$ in Eq. (\ref{eq:D18}), 
and because $N$ factors in the third and fourth terms 
in Eq. (\ref{eq:D22}) are cancelled out by $1/N$ factors of $D_q^2$ and $D_p^2$
in Eqs. (\ref{eq:D18}) and (\ref{eq:D19}).
This is the consequence of our choice of $c_n$ in Eq. (\ref{eq:D6}):
a different choice of the $N$ dependence of $c_n$ leads to $N$-dependent $R$. 
Furthermore $R$ is independent of $\beta$ because $\beta$ factor 
in the second term of Eq. (\ref{eq:F4}) is cancelled out by $1/\beta$ in Eq. (\ref{eq:D22}).

Figure \ref{fig3}(a) shows the $\tau$ dependence of $\mu$ ($= \langle W \rangle$)
for $c_o=0.0$ (solid curves), 0.5 (dashed curves) and 1.0 (chain curves). 
For $\tau \lesssim T_o\; (\simeq 6)$, we obtain 
$ \langle W \rangle > \Delta F$ ($=-0.5$) signaling the occurrence of the irreversibility.
At the same time, $\sigma$ ($= \sqrt{\langle (W-\langle W \rangle)^2} \rangle$) 
is rapidly increased for $\tau \lesssim T_o$, where 
fluctuation in $W$ much grows, as shown in Fig. \ref{fig3}(b). 
For $\tau= 2 m \pi/\Omega$ ($m=1,2,\cdot\cdot$) with $c_o=0.0$, 
$\sigma$ vanishes [Eq. (\ref{eq:F7})].
Figure \ref{fig3}(c) will be explained shortly.

Figure \ref{fig4} shows $3D$ plots of WDF of $P(W)$ as functions of $W$ and $\tau$ for $c_o=0.0$: 
result for $c_o=1.0$ is not so different from that of $c_o=0.0$ on first glance.
With decreasing $\tau$, the center of $P(W)$ moves to zero and
its width is considerably increased as Figs. \ref{fig3}(a) and \ref{fig3}(b) show.

\subsection{Jarzynski equality}

In this subsection, we consider the JE given by Eq. (\ref{eq:A1}). 
From Eqs. (\ref{eq:A1}) and (\ref{eq:F3})-(\ref{eq:F4}),
the JE is satisfied if the relation given by
\begin{eqnarray}
R &=& 
\phi - \frac{\beta \sigma^2}{2}
= \Delta F,
\label{eq:F8}
\end{eqnarray}
holds. Here $\Delta F$ denotes the free energy difference between the
two equilibrium systems with and without a force $g$ defined by \cite{Jarzynski04}
\begin{eqnarray}
\Delta F &=& F(g)-F(0) 
= - \frac{1}{\beta} \ln \frac{Z_S(g)}{Z_S(0)},
\label{eq:B5c}
\end{eqnarray}
with
\begin{eqnarray}
Z_S(g) &=& \frac{{\rm Tr}\:\{e^{-\beta[H_S(g)+H_B+H_I]} \}} 
{{\rm Tr}\:\{e^{-\beta H_B}\} }, 
\label{eq:B5d}
\end{eqnarray}
where
\begin{eqnarray}
H_S(g) &=& \frac{P^2}{2 M}
+\frac{M \Omega^2}{2}\left(Q- \frac{g}{M \Omega}\right)^2 - \frac{g^2}{2 M \Omega},
\label{eq:B5b}\\
H_B &=& \sum_{n=1}^N \left( \frac{p_n^2}{2 m}
+ \frac{m \omega_n^2 q_n^2}{2} \right),
\label{eq:B5f} \\
H_B + H_I &=& \sum_{n=1}^N \left[ \frac{p_n^2}{2 m}
+ \frac{m \omega_n^2 }{2}\left(q_n -\frac{c_n Q}{m \omega_n^2} \right)^2 \right],
\label{eq:B5e}
\end{eqnarray}
$Z_S(g)$ denoting the partition function of the system with $H_S(g)$ 
for a constant force of $f(t)=g$.
By using Eqs. (\ref{eq:B5c})-(\ref{eq:B5e}), we obtain 
\begin{eqnarray}
Z_S(g) &=& \left(\frac{2 \pi}{\beta \Omega} \right) e^{\beta g^2/2 M \Omega^2},
\end{eqnarray}
yielding
\begin{eqnarray}
\Delta F &=& -\frac{g^2}{2 M \Omega^2},
\label{eq:B5}
\end{eqnarray}
which is independent of the coupling $c_o$.

In what follows, we examine $\mu$, $\sigma^2$ and $R$ in the three limits of 
(1) no couplings $(c_o=0)$, (2) transient force $(\tau \rightarrow 0)$ and 
(3) quasi-static force $(\tau \rightarrow \infty)$.

\noindent
(1) In the limit of $c_o=0$, we obtain from 
Eqs. (\ref{eq:D22}) and (\ref{eq:D15})-(\ref{eq:D19}),
\begin{eqnarray}
\mu &=& - \frac{g^2}{2 M \Omega^2}
+ \frac{g^2 (1- \cos \Omega \tau)}{M \Omega^4 \tau^2}, \\
\sigma^2 &=& \frac{2 g^2 (1-\cos \Omega \tau)}{\beta M \Omega^4 \tau^2},
\label{eq:F7}
\end{eqnarray}
leading to
\begin{eqnarray}
R &=& 
- \frac{g^2}{2 M \Omega^2} = \Delta F,
\label{eq:F5}
\end{eqnarray} 
where $\Delta F$ is given by Eq. (\ref{eq:B5}).

\noindent
(2) In the limit of $\tau \rightarrow 0$, Eqs. (\ref{eq:D22}) and 
(\ref{eq:D15})-(\ref{eq:D19}) lead to
\begin{eqnarray}
\mu &=& 0,\\  
\sigma^2 &=& \frac{g^2}{\beta M \Omega^2}, 
\label{eq:G3}
\end{eqnarray}
where we employ the relations: $C_Q=-g$ and $C_P=D_q=D_p=0$.
A substitution of Eq. (\ref{eq:G3}) into Eq. (\ref{eq:F4}) leads to 
\begin{eqnarray}
R &=& -\frac{g^2}{2 M \Omega^2} = \Delta F.
\label{eq:G4}
\end{eqnarray}

\noindent
(3) In limit of $\tau \rightarrow \infty$, we obtain
\begin{eqnarray}
\mu &=& -g^2 \sum_{i=1}^2 \frac{b_i}{2 M a_i^2} = - \frac{g^2}{2 M \Omega^2}, \\
\sigma^2 &=& 0,
\label{eq:G1}
\end{eqnarray}
employing the relations: $\sum_{i=1}^2 (b_i/a_i^2)=1/\Omega^2$ 
and $C_Q=C_P=D_q=D_p=0$.
Equations (\ref{eq:F4}) and (\ref{eq:G1}) lead to
\begin{eqnarray}
R &=& - \frac{g^2}{2 M \Omega^2}=\Delta F.
\label{eq:G2}
\end{eqnarray}
Equations (\ref{eq:F5}), (\ref{eq:G4}), and (\ref{eq:G2}) imply that the JE holds
in the three limits of (1) $c_o=0$, (2) $\tau \rightarrow 0$ 
and (3) $\tau \rightarrow \infty$.

Figure \ref{fig3}(c) shows that the JE is numerically verified
for $10^{-1} \leq \tau \leq 10^2$ with $c_o=0.0$, 0.5 and 1.0.
The JE is valid even when we adopt other sets of model parameters.
It is surprising that complicated expressions of $\mu$ $(=\phi)$ and $\sigma^2$
given by Eqs. (\ref{eq:D15}) and (\ref{eq:D22}), respectively, 
satisfy the JE given by Eq. (\ref{eq:F8}).
Although the validity of the JE is confirmed by numerical calculations,
we have not succeeded in its analytical proof 
except for the three cases of $c_o=0$, $\tau \rightarrow 0$ and $\tau \rightarrow \infty$. 

\section{Discussion}

\subsection{Canonical average over initial equilibrium state}

It should be stressed that the canonical average in Eq. (\ref{eq:H3b}) must be performed
over the total Hamiltonian $H$ $(=H_S+H_B+H_I)$ in the initial equilibrium state \cite{Talkner}.
If the average in Eq. (\ref{eq:H3b}) is performed over the Hamiltonian
of the uncoupled state $(H_S+H_B)$ in place of $H$, we obtain a wrong result.
Figures \ref{fig5}(a) and (b) show $\bar{E}_S(t)$ $(=\langle E_S(t) \rangle_{00})$ 
with no forces ($f=0.0$) and a ramp force
of $\tau=100$, respectively, with $c_o=1$ 
when the average is performed over the initial uncoupled state of $(H_S+H_B)$,
\begin{eqnarray}
\langle O \rangle_{00} 
&\equiv & \frac{{\rm Tr} \:\{e ^{-\beta [H_S(0)+H_B(0)]} \:O \} }
{{\rm Tr} \:e ^{-\beta [H_S(0)+H_B(0)]}},
\label{eq:H6}
\end{eqnarray}
where $O$ stands for an operator.
Results in Figs. \ref{fig5}(a) and (b) are quite different from the
corresponding ones averaged over $H$ which have been shown in Figs. \ref{fig2}(b) and (d).
In particular, the irregular energy exchange between the system and bath occurs even 
when $f(t)=0.0$ in Fig. \ref{fig5}(a), while the initial serene state 
persists in Figs. \ref{fig2}(b).
Figure \ref{fig5}(a) denotes the result of the case where
the system-bath coupling is suddenly added at $t = 0.0$ to
the uncoupled system in equilibrium state at $t < 0.0$.  
The perturbation of the added coupling induces the irregular energy exchange between
the system and bath which does not dissipate for $t \geq 0.0$.

If the canonical average in Eq. (\ref{eq:F9}) is performed over $H_S+H_B$, we obtain
\begin{eqnarray}
\langle e^{-\beta W_0}\rangle_{00} 
&=& e^{-\beta(\phi-\beta \sigma'\:^2/2)},
\end{eqnarray}
with
\begin{eqnarray}
\sigma'\:^2 &=& \frac{1}{\beta}\left[ \frac{C_Q^2}{M \Omega^2} +M C_P^2
+ \frac{N D_q^2}{m \omega_o^2} + m N D_p^2 \right],
\label{eq:H7}
\end{eqnarray}
where $\phi$ is given by Eq. (\ref{eq:D14b}). 
Because $\sigma'\:^2$ is different from $\sigma^2$ in Eq. (\ref{eq:D22}),  
it wrongly leads to a violation of the JE: 
$R= \phi-\beta \sigma'\:^2/2 \neq \phi-\beta \sigma^2/2=\Delta F$.
The related discussion will be given also in Sec. III.D.

\subsection{A two-step ramp force}
Besides a ramp force given by Eq. (\ref{eq:B4}), we have employed
a two-step ramp force given by
\begin{eqnarray}
f(t) &=& \left\{ \begin{array}{ll}
0
\quad & \mbox{for $t < 0 $}, \\ 
g \left( \frac{h \:t}{\tau_m} \right)
\quad & \mbox{for $0 \leq t < \tau_m $}, \\ 
g \left[\frac{(1-h)t+(h \:\tau-\tau_m)}{(\tau-\tau_m)} \right]
\quad & \mbox{for $\tau_m \leq t < \tau $}, \\
g
\quad & \mbox{for $t \geq \tau $},
\end{array} \right. 
\label{eq:K1}
\end{eqnarray}
where $h$ stands for a magnitude at a middle time of $\tau_m$ $(< \tau)$.
Figure \ref{fig6} (a)and (b) show $\mu$ and $\sigma$, respectively,
as a function of $\tau$ when a two-step ramp input $f(t)$ given 
by Eq. (\ref{eq:K1}) with $g=1.0$, $h=1.5$ and $\tau_m=\tau/2$ is applied
[$f(t)$ is shown in the inset of Fig. \ref{fig6}(b)].
The input force $f(t)$ first linearly increases to $1.5 g$ at $t=\tau_m$ and
then it linearly decreases to the final value of $g$ at $t \geq \tau$.
The $\tau$ dependences of $\mu$ and $\sigma$ shown in Fig. \ref{fig6} are rather different 
from those for a single-step ramp input shown in Fig. \ref{fig3}.
In particular, magnitudes of $\mu$ and $\sigma$ have resonance-like peaks at $\tau \sim T_o$.
Nevertheless the JE holds also for the two-step ramp force.

\subsection{Baths with multiple $\omega$ and infinite $N$}
Our study in the preceding section has been made for the CL model 
with the single-$\omega$ bath, which may be extended to multiple-$\omega$ bath.
The Green's function given by Eq. (\ref{eq:D4}) may be generally expressed by
\begin{eqnarray}
\hat{G}(s) &=& \sum_{i=1}^{N+1} \frac{b_i}{(s^2+\tilde{\omega}_i^2)},
\label{eq:E10}
\end{eqnarray}
where $\tilde{\omega}_i$ denotes the normal-mode frequency of the coupled system-plus-bath
and $b_i$ is expressed in terms of the corresponding eigenfunction \cite{Dhar07}.

The inverse Laplace transformation leads to
\begin{eqnarray}
G(t) &=& \sum_{i=1}^{N+1} \frac{b_i \sin \tilde{\omega}_i t}{\tilde{\omega}_i}.
\label{eq:E11}
\end{eqnarray} 
The Green's function given by Eq. (\ref{eq:E10}) or (\ref{eq:E11})
has the same structure as that for the single-$\omega$ bath given by
Eq. (\ref{eq:D8}).
Calculations of $Q(t)$, $W_0$ and $P(W)$ may be formally performed
in the same way as was made in Sec. II.  
Then properties of the CL model with finite-$N$ multiple-$\omega$ bath are essentially
the same as those with single-$\omega$ bath.

On the other hand, in the limit of $N \rightarrow \infty$, the summation over $n$ 
in the Green's function of Eq. (\ref{eq:D4}) is converted to integral
over a continuous spectrum and it may be expressed by
\begin{eqnarray}
\hat{G}(s) &=& \left[ s^2+\Omega^2 + \frac{s^2 c_o^2}{M m} 
\int\frac{ D(\omega)}{w^2 (s^2+\omega^2)} \:d\omega \right]^{-1},
\label{eq:K2}
\end{eqnarray}
where $D(\omega)$ denotes the density of state,
\begin{eqnarray}
D(\omega) &=& \frac{1}{N} \sum_{n=1}^N \:\delta(\omega -\omega_n).
\label{eq:K3}
\end{eqnarray}
When we assume the Debye-type density of states: $D(\omega) = a \:\omega^2$
($a$: constant), $\hat{G}(s)$ is given by
\begin{eqnarray}
\hat{G}(s) &=& \frac{1}{(s+c_1)(s+c_2)},
\label{eq:K4}
\end{eqnarray}
with 
\begin{eqnarray}
c_{1,2} &=& \pm i \sqrt{\Omega^2-\left( \frac{\pi a c_o^2}{4 M m} \right)^2}
+\left( \frac{\pi a c_o^2}{4 M m} \right).
\label{eq:K5}
\end{eqnarray}
Because of real parts in $c_1$ and $c_2$, the inverse Laplace transformation of $\hat{G}(s)$ 
in Eq. (\ref{eq:K4}) yields dissipative $G(t)$ which vanishes at $t \rightarrow \infty$. 
For dissipation it is necessary that the frequencies $\{ \omega_n \}$ have
a continuous spectrum in the limit of $N \rightarrow \infty$ \cite{Dhar07}.
With a discrete spectrum for finite $N$, however, the Green's function $G(t)$ 
in Eq. (\ref{eq:D4}) is non-dissipative and not vanishing 
in the limit of $t \rightarrow \infty$.

\subsection{The generalized Langevin approach}
In the conventional approach to the CL model, we derive the Langevin equation given by
\begin{eqnarray}
M \ddot{Q} &=& -M \Omega^2 Q -\int_0^{t} \gamma(t-t')\dot{Q}(t') \:dt'
+\zeta'(t)+f(t),
\label{eq:E1}
\end{eqnarray}
with 
\begin{eqnarray}
\zeta'(t) &=& \zeta(t) -\gamma(t)Q(0), 
\label{eq:E2b} \\
\gamma(t) &=& \sum_{n=1}^N \left(\frac{c_n^2}{m \omega_n^2}\right) \cos \omega_n t, 
\label{eq:E2}\\
\zeta (t)&=& \sum_{n=1}^N c_n \left[ q_n(0) \cos \omega_n t 
+\left( \frac{p_n(0)}{m \omega_n}\right) \sin \omega_n t \right].
\label{eq:E3}
\end{eqnarray}
after obtaining a formal solution of $q_{n}(0)$ from Eq. (\ref{eq:B3}) 
and substituting it into Eq. (\ref{eq:B2}) \cite{Caldeira81,Caldeira83}.
Equations (\ref{eq:E1})-(\ref{eq:E3}) express the non-Markovian Langevin equation
with colored noise. 

When we adopt the single-$\omega$ bath given by Eq. (\ref{eq:D5a}), 
$\gamma(t)$ and $\zeta(t)$ are given by
\begin{eqnarray}
\gamma(t) &=& \left( \frac{c_o^2}{m \omega_o^2}\right) \cos \omega_o t, 
\label{eq:E5}\\
\zeta (t)&=& \frac{c_o}{\sqrt{N}} \left[
\cos \omega_o t \sum_{n=1}^N  q_{n0}
+ \frac{\sin \omega_o t}{m \omega_o} \sum_{n=1}^N p_{n0} \right].
\label{eq:E6}
\end{eqnarray}
By using the Laplace transformation yielding
\begin{eqnarray}
\hat{\gamma}(s) &=& 
\frac{c_o^2 \:s}{m \omega_o^2 (s^2+\omega_o^2)},
\label{eq:E7} \\
\hat{\zeta}(s)&=& \frac{c_o}{\sqrt{N}} \left[
\frac{s}{s^2+\omega_o^2} \sum_{n=1}^N  q_{n0}
+ \frac{1}{m (s^2+\omega_o^2)} \sum_{n=1}^N p_{n0} \right],
\label{eq:E8b}
\end{eqnarray}
we obtain an equation for $\hat{Q}(s)$ which is exactly the same
as Eqs. (\ref{eq:D6}) and (\ref{eq:D7}).

It has been shown that the JE is satisfied in the non-Markovian Langevin model
with colored noise (generalized Langevin model)
\cite{Zamponi05,Mai07,Speck07,Ohkuma07},
which is different from our Langevin equation given by 
Eqs. (\ref{eq:E1}), (\ref{eq:E5}) and (\ref{eq:E6})
in the two points: (a) the second term of Eq. (\ref{eq:E2b}) includes an additional term of
$- \gamma(t) Q(0)$ which is missing in the conventional generalized Langevin model, and 
(b) the memory kernel given by Eq. (\ref{eq:E5}) is oscillating and non-dissipative while
that in the generalized Langevin model is dissipative.
In the literature ({\it e.g.} Ref. \cite{Ford87}), 
the additional term of $- \gamma(t) Q(0)$ is discarded and
the fluctuation-dissipation relation is given by 
\begin{eqnarray}
\langle \zeta(t) \:\zeta(t') \:\rangle_{00}= k_B T \gamma(t-t'),
\label{eq:E9}
\end{eqnarray}
which is derived from the equi-partition relations:
$\langle q_{n0} \:q_{\ell 0}\rangle_{00}=(k_B T/m \omega_n^2) \delta_{n \ell}$
and $\langle p_{n0} \:p_{\ell 0}\rangle_{00}=(k_B T m) \delta_{n \ell}$,
the bracket $\langle \cdot \rangle_{00}$ denoting
the canonical average over the uncoupled initial state $H_S+H_B$ [Eq. (\ref{eq:H6})].
If the additional term is included, we obtain 
\begin{eqnarray}
\langle \zeta'(t) \:\zeta'(t') \rangle_{00} 
&=& k_B T \left[\gamma(t-t') +\frac{\gamma(t) \gamma(t')}{M \Omega^2}
\right],
\end{eqnarray}
which is different from Eq. (\ref{eq:E9}).
It is noted, however, that when employing Eqs. (\ref{eq:H2a})-(\ref{eq:H2b}) valid for
equilibrium initial states of coupled Hamiltonian $H$ ($=H_S+H_B+H_I$),
we obtain the desired fluctuation-dissipation relation,
\begin{eqnarray}
\langle \zeta'(t) \zeta'(t') \rangle_0 &=& k_B T \gamma(t-t').
\label{eq:E4}
\end{eqnarray}
Then the ostensible inconsistency of the item (a) may be resolved.
As for the item (b),
the importance of the ergodicity is emphasized in Refs. \cite{Speck07,Ohkuma07}
from a study on the JE for the generalized Langevin model.
Non-ergodic solutions of the non-dissipative generalized Langevin 
equation have been discussed in Refs. \cite{Dhar07,Bao05,Plyukhin11}.
Our non-diffusive memory kernel 
yields non-ergodic solutions for the Langevin equation given by Eqs. (\ref{eq:E1}), 
(\ref{eq:E5}) and (\ref{eq:E6}).
It is noted that the JE holds in our calculation even if the condition of the ergodicity 
is not satisfied, in contrast with Refs. \cite{Speck07,Ohkuma07}.

Before closing Sec. III, it is necessary to mention that Ref. \cite{Sung05c} has 
studied the validity condition of the JE for a general classical dynamical system
with any time-dependent external force $f(t)$. It is shown in \cite{Sung05c} that
the JE holds for a classical system during a transition process in
which the value of a parameter $f$ in the system Hamiltonian switches
from $f_0$ to $f_1$ in time $\tau$, as long as the phase space extension 
of the system $\Omega_0^{eq}$ at the initial equilibrium phase space 
with $f = f_0$ is the same as the equilibrium phase space $\Omega_1^{eq}$ 
with $f = f_1$. 
It is noted that the general validity condition reported in Ref.\cite{Sung05c}
is satisfied for our system even though it is a non-ergodic one.

\section{Conclusion}
We have studied the response to an applied force of small open oscillator system 
described by the exactly solvable CL model with the non-dissipative single-$\omega$ bath. 
Although the model adopted in our study seems a pedagogical toy model,
it is expected not to be unrealistic because non-dissipative properties are realized 
in small systems coupled to finite baths \cite{Plyukhin01,Hasegawa11}.
We have obtained exact expressions for position, momentum and energy of the system
whose canonical averages have been analytically performed over initial equilibrium states.
Our calculations of system energy and work have shown the following:

\noindent
(i) the energy of the system strongly coupled to finite bath is fluctuating 
but non-dissipative in general,
and

\noindent
(ii) the JE is valid in non-dissipative non-ergodic systems. 

\noindent
The item (i) supports direct simulations for open systems coupled to finite baths 
\cite{Plyukhin01,Hasegawa11} although it is contrast to the result 
showing the dissipation for $N \gtrsim 10-20$ \cite{Rosa08}.
The item (ii) is consistent with Jarzynski's proof for an arbitrary classical open systems
\cite{Jarzynski04}.
Our study is complementary to the previous studies for dissipative
oscillator systems with the use of
the Markovian \cite{Douarche05,Douarche06,Joubaud07,Joubaud07b}
and non-Markovian Langevin models \cite{Zamponi05,Mai07,Speck07,Ohkuma07}, 
Fokker-Planck equation \cite{Chaudhury08}
and Hamiltonian models \cite{Jarzynski06,Jarzynski08,Dhar05,Chakrabarti08,Hijar10}.

Although the items (i) and (ii) hold for open systems
described by the CL \cite{Caldeira81,Caldeira83} and Ford-Kac models \cite{Ford87},
it is not certain whether they are valid for any non-dissipative non-ergodic open system.
In this respect, it would be interesting to examine a work in the $(N_S+N_B)$ model
for a classical $N_S$-body system coupled to an $N_B$-body bath \cite{Hasegawa11}.
The $(N_S+N_B)$ model clarifies some interesting issues such as the $N_S$-dependent 
non-Gaussian energy distribution of the system \cite{Hasegawa11} 
which has been not realized in previous studies for CL-type models 
with $N_S=1$ \cite{Caldeira81,Caldeira83,Ford87}.
Such a calculation is in progress and will be reported in a separate paper.

\appendix*

\begin{acknowledgments}
The author expresses his sincere thanks to Prof. Peter Talkner and Prof. Juyeon Yi 
for pointing out a mistake in the first version of the manuscript. 
This work is partly supported by
a Grant-in-Aid for Scientific Research from 
Ministry of Education, Culture, Sports, Science and Technology of Japan.  
\end{acknowledgments}


\newpage
\begin{figure}
\begin{center}
\end{center}
\caption{
(Color online) 
The time dependence of the averaged position $\bar{Q}$ (solid curves) 
and momentum $\bar{P}$ (dashed curves);
(a) $\tau=100.0$, $c_o=0.0$, (b) $\tau=100.0$, $c_o=1.0$, 
(c) $\tau=10.0$, $c_o=0.0$, (d) $\tau=10.0$, $c_o=1.0$,
(e) $\tau=5.0$, $c_o=0.0$, (f) $\tau=5.0$, $c_o=1.0$,
(g) $\tau=0.0$, $c_o=0.0$, and (h) $\tau=0.0$, $c_o=1.0$.
}
\label{fig1}
\end{figure}

\begin{figure}
\begin{center}
\end{center}
\caption{
(Color online) 
The time dependence of the averaged system energy $\bar{E}_S$; for no forces ($f=0$)
with (a) $c_o=0.0$, (b) $c_o=1.0$; for the ramp forces with
(c) $\tau=100.0$, $c_o=0.0$, (d) $\tau=100.0$, $c_o=1.0$, 
(e) $\tau=10.0$, $c_o=0.0$, (f) $\tau=10.0$, $c_o=1.0$, 
(g) $\tau=5.0$, $c_o=0.0$, (h) $\tau=5.0$, $c_o=1.0$,
(i) $\tau=0.0$, $c_o=0.0$, and (j) $\tau=0.0$, $c_o=1.0$.
}
\label{fig2}
\end{figure} 

\begin{figure}
\begin{center}
\end{center}
\caption{
(Color online) 
The $\tau$ dependence of (a) $\mu$ (=$\langle W \rangle$), 
(b) $\sigma$ ($= \sqrt{\langle (W-\langle W \rangle)^2 \rangle }$) and 
(c) $R$ $(=- \beta^{-1} \ln \langle e^{-\beta W} \rangle)$
for $c_o=0.0$ (solid curves), 0.5 (dashed curves) and 1.0 (chain curves);
arrows along the right ordinates in (a) and (c) express $\Delta F$ ($=- 0.5$).
In (c) $R=\Delta F$ for $c_o=0.0$, 0.5 and 1.0.
}
\label{fig3}
\end{figure}

\begin{figure}
\begin{center}
\end{center}
\caption{
(Color online) 
$3D$ plots of $P(W)$ as functions of $W$ and $\tau$ for $c_o=0.0$,
the ordinate of (b) being enlarged compared to that of (a).
}
\label{fig4}
\end{figure}

\begin{figure}
\begin{center}
\end{center}
\caption{
(Color online) 
The time dependence of $\bar{E}_S(t)$ with (a) no forces $(f=0)$ and (b) a ramp force of $\tau=100.0$ 
with $c_o=1.0$ when the average is performed over initial uncoupled state of $H_S+H_B$:
(a) and (b) should be compared to Figs. \ref{fig2}(b) and (d), respectively,
which are averaged over the initial coupled state of $H_S+H_B+H_I$ (see text).
}
\label{fig5}
\end{figure} 

\begin{figure}
\begin{center}
\end{center}
\caption{
(Color online) 
The $\tau$ dependence of (a) $\mu$ (=$\langle W \rangle$) and 
(b) $\sigma$ ($= \sqrt{\langle (W-\langle W \rangle)^2 \rangle }$) 
for a two-step ramp force $f(t)$ given by Eq. (\ref{eq:K1}) 
with $g=1.0$, $h=1.5$ and $\tau_m=\tau/2$ [see the inset of (b)] 
with $c_o=0.0$ (solid curves),
0.5 (dashed curves) and 1.0 (chain curves);
an arrow along the right ordinate in (a) expresses $\Delta F$ ($=- 0.5$).
}
\label{fig6}
\end{figure}

\end{document}